\documentstyle[12pt]{article}
\input psfig.sty
\begin{document}

\title{The Exact Solution to\\
the Schr\"{o}dinger Equation with the Octic Potential}

\author{Shi-Hai Dong\thanks{Electronic address: DONGSH@HPTC5. IHEP. AC. CN}\\
{\footnotesize Institute of High Energy Physics, P. O. Box 918(4), 
Beijing 100039, People's Republic of China}
\\
Zhong-Qi Ma\\
{\footnotesize China Center for Advanced Science and Technology
(World Laboratory), P. O. Box 8730, Beijing 100080}\\
{\footnotesize  and Institute of High Energy Physics, P. O. Box 918(4), 
Beijing 100039, People's Republic of China}}

\date{}

\maketitle

\begin{abstract}

The Schr\"{o}dinger equation with the central potential is first studied 
in the arbitrary dimensional spaces and obtained an analogy of the 
two-dimensional Schr\"{o}dinger equation for the radial wave function
through a simple transformation. As an example, 
applying an ${\it ansatz}$ to the eigenfunctions, 
we then arrive at an exact closed form solution to 
the modified two-dimensional Schr\"{o}dinger equation with the octic 
potential, $V(r)=ar^2-br^4+cr^6-dr^4+er^{10}$. 


\end{abstract}

\newpage

\begin{center}
{\large 1. Introduction}\\
\end {center}

It is well known that the general framework of the nonrelativistic quantum
mechanics is by now well understood[1, 2], however,  
whose predictions have been
carefully proved against observations[3]. It is of importance to know
whether some familiar problems are a particular case of a more general scheme. 
On behalf of this purpose, it is worthwhile to study the 
Schr\"{o}dinger equation in the arbitrary dimensional spaces. This topic has 
attracted much more attention to many authors[4-8]. With respect to the 
arbitrary dimensional Schr\"{o}dinger equation, it is readily to arrive at 
a simple analogy of the two-dimensional Schr\"{o}dinger 
equation for the radial wave function through a simple transformation.

On the other hand, the exact solutions to the fundamental
dynamical equations play an important
role in physics. As we know, the exact solutions
to the Schr\"{o}dinger equation
are possible only for several
potentials and some approximation methods
are frequently used to arrive
at the solutions. Recently, the study of  higher
order anharmonic potentials have been much more desirable
to physicists and
mathematicians[10-12], who want to understand
a few newly discovered phenomena
(for instance, structural phase transitions[10], polaron
formation in solids[11] and the
concept of false vacuo in filed theory[12]) in the
different fields of physics. 
Unfortunately, in these anharmonic potentials, 
not much work has been
carried out on the octic potential
except for some simpler study[13] by an
${\it ansatz}$ to the eigenfunctions in the 
three-dimensional spaces. With the wide interest in the
lower-dimensional field theory recently, however, 
it seems reasonable to
study the two-dimensional Schr\"{o}dinger equation 
with the octic potential. 
We has succeeded in dealing with
the Schr\"{o}dinger equation with some anharmonic potentials 
by this ${\it ansatz}$[15-17]. 
Consequently, we attempt to study 
the two-dimensional Schr\"{o}dinger equation 
with the octic potential, to our knowledge,  which is not 
appeared in the literature. 
The purpose of this paper is to demonstrate the modified 
Schr\"{o}dinger equation in the arbitrary dimensional spaces and 
give a concrete application to the two-dimensional 
Schr\"{o}dinger equation with the octic potential.

The paper is organized as follows. Section 2 studies 
the Schr\"{o}dinger equation with the central potential  
in the arbitrary dimensional spaces and obtains an analogy of the 
two-dimensional Schr\"{o}dinger equation for the radial wave function
through a simple transformation. In Sec. 3, as an example, 
applying an ${\it ansatz}$ to the eigenfunctions, 
we obtain an exact closed form solution to 
the modified two-dimensional Schr\"{o}dinger equation 
with the octic potential. 

\vskip 1cm
\begin{center}
{\large 2. The modified Schr\"{o}dinger equation}
\end{center}

Throughout this paper the natural unit $\hbar=1$
and $\mu=1/2$ are employed. Following the Refs. [7, 8], 
in the $N$ dimensional Hilbert spaces, the radial 
wave function $\psi(r)$ for the 
Schr\"{o}dinger equation for the
stationary states can be written as 
$$\left[\frac{d^2}{dr^2}+\frac{(N-1)}{r}\frac{d}{dr}
+(E-V(r))-\frac{\ell(\ell+N-2)}{r^2}\right]\psi(r)=0, \eqno(1)$$

\noindent
where $\ell$ denotes the angular momentum quantum number. 
In order to make the coefficient of the first derivative vanish, we may
furthermore define a new radial wave 
function $R(r)$ by means of the equation[9], 
$$\psi(r)\equiv r^{\rho}R(r), \eqno(2a)$$

\noindent
where $\rho$ is an unknown parameter and will be given in the following. 
Substituting Eq. (2a) into Eq. (1), we will arrive at
an algebraic equation containing the parameter $\rho$ as
$$2\rho+(N-1)=0. $$

\noindent
Consequently, the Eq. (2a) will be read as
$$\psi\equiv r^{\frac{-(N-1)}{2}}R(r), \eqno(2b)$$ 

\noindent
which will lead to the radial wave function $R(r)$ satisfying
$$\left\{\frac{d^2}{dr^2}-\left[\ell(\ell+N-2)+\frac{1}{4}(N-1)(N-3)
+(E-V(r))\right]\frac{1}{r^2}\right\}R(r)=0. \eqno(3)$$

\noindent
Through a simple deformation, 
$$\ell(\ell+N-2)+\frac{1}{4}(N-1)(N-3)
=\left[\ell+\frac{1}{2}(N-2)\right]^{2}-\frac{1}{4}, $$

\noindent
we may introduce a parameter
$$\eta\equiv \ell+\frac{1}{2}(N-2), \eqno(4)$$

\noindent
so that the Eq. (3) will be written as
$$\left[\frac{d^2}{dr^2}+(E-V(r))-\frac{(\eta^{2}-1/4)}{r^2}\right]R(r)=0, 
\eqno(5)$$

\noindent
which is our desired result. In other words, we have modified the
Schr\"{o}dinger equation in the arbitrary dimensional spaces into a simple
analogy of the two-dimensional radial Schr\"{o}dinger equation after
introducing a parameter $\eta$ given in Eq. (4), 
which relies on a linear combination
between $N$ and the angular momentum quantum number $\ell$. 
As mentioned above, we want to solve this modified Schr\"{o}dinger equation
with the octic potential in two dimensions applying an ${\it ansatz}$ to the
eigenfunctions in the next section. 

\vskip 1cm
\begin{center}
{\large 3. An ${\it ansatz}$ to the eigenfunctions}
\end{center}

Consider the two-dimensional 
Schr\"{o}dinger equation with a potential $V(r)$
that depends only on the distance $r$ from the origin
$$H\psi(r, \varphi) =\left(
\displaystyle {1 \over r} \displaystyle {\partial \over \partial r} 
r \displaystyle {\partial \over \partial r} + 
\displaystyle {1 \over r^{2}} \displaystyle {\partial^{2} \over 
\partial \varphi^{2} } \right)\psi(r, \varphi) +V(r) \psi(r, \varphi) 
=E \psi(r, \varphi), \eqno(6)$$ 

\noindent
where here and hereafter the potential
$$V(r)=ar^2-br^4+cr^6-dr^8+er^{10}, ~~ d<0. \eqno(7)$$

\noindent
Due to the symmetry of the potential, let
$$\psi(r, \varphi)=r^{-1/2} R_{m}(r) e^{ \pm im \varphi}, 
~~~~~m=0, 1, 2, \ldots. \eqno (8) $$  

\noindent
It is easy to find from Eq. (5) that 
the radial wave function $R_{m}(r)$ satisfies the following 
radial equation
$$\displaystyle {d^{2} R_{m}(r) \over dr^{2} }
+\left[E-V(r)-\displaystyle {m^{2}-1/4 \over r^{2}} \right] R_{m}(r)=0, 
\eqno (9) $$

\noindent
where the parameter $\lambda=m=\ell, N=2$; $m$ and $E$ denote the
angular momentum number and energy, respectively. 
For the solution of Eq. (9), we make an ${\it ansatz}$[13-15] 
for the ground state
$$R_{m0}(r)=\exp[p_{m0}(r)], \eqno(10)$$
 
\noindent
where
$$p_{m0}(r)=\frac{1}{2} \alpha r^2
-\frac{1}{4}\beta r^4+\frac{1}{6}\tau r^6+\kappa \ln r. \eqno(11)$$

\noindent
After calculating, we arrive at the following equation
$$\displaystyle {d^{2} R_{m0}(r) \over dr^{2}}
-\left[\displaystyle{d^{2} p_{m0}(r) \over dr^{2}}
+\left(\displaystyle {dp_{m0}(r) \over dr}\right)^2\right]R_{m0}(r)=0. 
\eqno(12)$$

\noindent
Compare Eq. (12) with Eq. (9) as before and
obtain the following set of equations
$$\kappa (\kappa-1)=(m+1/2)(m-1/2), ~~~\tau^2=e, \eqno(13a)$$
$$\beta^2+2\alpha\tau=c, ~~~2\beta\tau=d, \eqno(13b)$$
$$\alpha^2-2\beta\kappa-3\beta=a, \eqno(13c)$$
$$5\tau+2\tau\kappa-2\alpha\beta=-b, \eqno(13d)$$
$$E=-\alpha(1+2\kappa). \eqno(13e)$$

\noindent
It is not difficult to obtain the
values of parameters $\tau$ and $\kappa$
from the Eq. (13a) written as
$$\tau=\pm \sqrt{e}, ~~~\kappa=-m+1/2~~~{\rm or}~~m+1/2. \eqno(14)$$

\noindent
In order to retain the well-behaved solution
at the origin and at infinity, we choose positive sign in $\tau$ and
$\kappa$ as $m+1/2$. According to these choices, 
the Eq. (13b) will give the other parameter values as
$$\beta=-\frac{d}{2\sqrt{e}}, ~~~\alpha=\frac{d^2-4ce}{8e\sqrt{e}}. \eqno(15)$$

\noindent
Besides, it is readily to obtain from the Eqs. (13c) and (13d) that
$$a=\frac{d^4-8ced^2+16c^2e^2+64de^2\sqrt{e}(\kappa+3/2)}{64e^3}, \eqno(16a)$$
$$b=\frac{8e^2\sqrt{e}(5+2\kappa)-d(d^2-4ec)}{8e^2}, \eqno(16b)$$

\noindent
which are the constraints on the parameters of the octic potential. 

\noindent
The eigenvalue $E$, however, will be given by Eq. (13e) as
$$E=-\frac{(1+2\kappa)(d^2-4ce)}{8e\sqrt{e}}. \eqno(17)$$

\noindent
The corresponding eigenfunctions Eq. (10) can now be read as
$$R_{m0}=N_{0}r^{\kappa}\exp\left[\frac{1}{2}\alpha^2-\frac{1}{4}\beta r^4+
\frac{1}{6}\tau r^6\right], \eqno(18)$$

\noindent
where $N_{0}$ is the normalized constant and here and hereafter the parameters
$\alpha, \beta$ and $\kappa$ are the same as the values given above. 
As a matter of fact, the normalized 
constant $N_{0}$ can be
calculated in principle from the normalized relation
$$\int_{0}^{\infty}|R_{m0}|^2dr=1, \eqno(19)$$

\noindent
which implies that
$$N_{0}=\left[\frac{1}{\omega}\right]^{1/2}, \eqno(20)$$

\noindent
where
$$\omega \equiv \int_{0}^{\infty}r^{2\kappa}\exp\left[\alpha r^2
-\frac{1}{2}\beta r^4+\frac{1}{3}\tau r^6\right] dr. \eqno(21)$$

\noindent
In this case, however, the normalization of the eigenfunctions becomes a
very difficult task. 
Considering the values of the
parameters of the potential, we fix them as follows. 
The value of parameter $c, d, e$ are first fixed, 
for example $c=1. 0, d=-2$ and $e=4. 0$, 
the value of the parameters $a$ and $b$ are given by
Eq. (11) for $m=0$. By this way, 
the parameters turn out to
$a=-1. 96, b=11. 8, c=4. 0, d=-2. 0, e=4. 0$ 
and $\alpha=-0. 188, \beta=0. 5, \tau=-2$. 
The ground state
energy corresponding to these
values is obtained as $E=0. 375$. 
Actually, when we study the property
of the ground state, as we know, the unnormalized
radial wave function will not
affect the main features of the wave function. 
We have plotted the unnormalized
radial wave function $R_{00}(r)$
in fig. 1 for the ground 
state. 

To summarize,  we first deal with the Schr\"{o}dinger 
equation with the central potential in the 
arbitrary dimensional spaces and obtain an 
analogy of the two-dimensional Schr\"{o}dinger equation for the 
radial wave function through a simple transformation.  
As an example, we obtain an exact closed solution to 
the Schr\"{o}dinger equation with 
the octic potential using 
a simpler ${\it ansatz}$ and simultaneously 
two constrains on the 
parameters of the potential are arrived at from the 
compared equation. 
The other studies to the Schr\"{o}dinger 
equation with the related anharmonic 
potential in two dimensions are in progress.

\vspace{10mm}
{\bf Acknowledgments}. This work was supported by the National
Natural Science Foundation of China and Grant No. LWTZ-1298 from 
the Chinese Academy of Sciences. 

%\newpage

%\end{document}

\newpage
\vskip 1cm
\begin{figure}
\begin{center}
%\leavevmode
 % \epsfxsize 0. 7\textwidth
 % \epsfbox{boun3. eps}
%\end{center}
\mbox{\psfig{figure=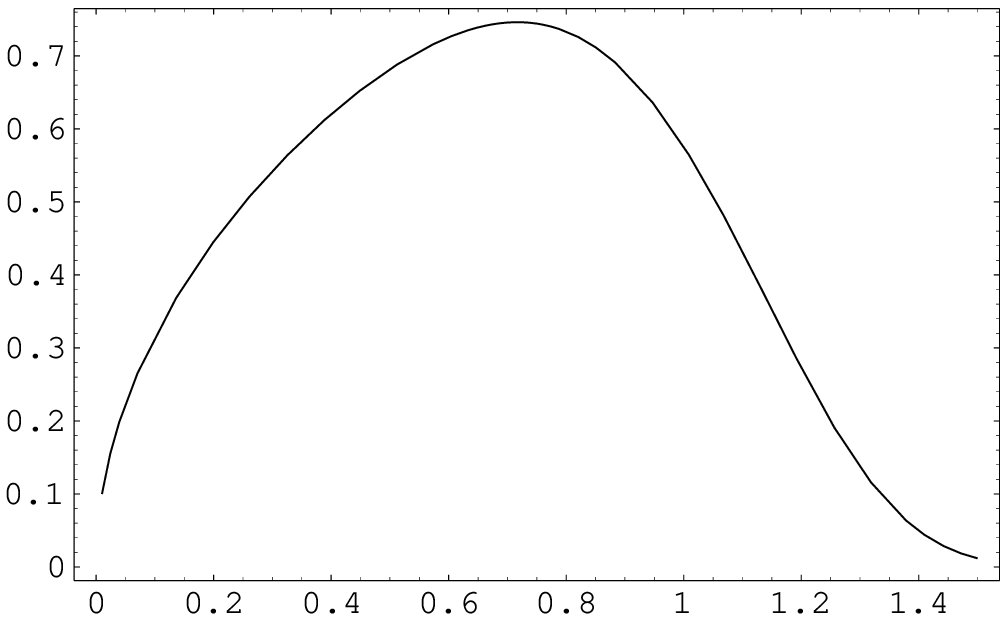,height=8cm,width=10cm}}
\end{center}
\caption{The ground state wave functions $R_{00}(r)$ as
a function of $r$ for the
potential (2) with the values
$a=-1. 96, b=11. 8, c=1. 0, d=-2. 0, e=4. 0$. 
The $x$-axis denotes the
variable $r$ and the $y$-axis denotes the values of wave functions. } 
\end{figure}

\end{document}